\documentclass[a4paper,11pt]{article}  
  
\usepackage{amsfonts,amssymb}  
\usepackage{amsmath}

\def\G{\Gamma}

\def\GG{{\rm I}\!\Gamma}

\begin{document}

\rightline{MPI-PhT/2002-79}
\vskip 3 truecm
\Large
\bf
\centerline{Slavnov-Taylor Parameterization for}
\centerline{the Quantum Restoration of BRST Symmetries}
\centerline{in Anomaly-free Gauge Theories}

\normalsize
\rm
\vskip 1.3 truecm
\large
\centerline{Andrea Quadri\footnote{E-mail address: {\tt quadri@mppmu.mpg.de}}}

\vskip 0.3 truecm
\normalsize
\centerline{Max-Planck-Institut f\"ur Physik}
\centerline{(Werner-Heisenberg-Institut)}
\centerline{F\"ohringer Ring, 6 - D80805 M\"unchen, Germany}

\vskip 1.3  truecm
\normalsize
\bf
\centerline{Abstract}

\rm
\begin{quotation}
It is shown that the problem of the recursive 
restoration of the Slavnov-Taylor
(ST) identities at the quantum level for anomaly-free gauge theories
is equivalent to the problem of parameterizing the local approximation
to the quantum effective
action in terms of ST functionals,
associated with the
cohomology classes of the classical linearized ST operator ${\cal S}_0$.
The ST functionals of dimension $\leq 4$ correspond to the invariant
counterterms, those of dimension $>4$ generate the non-symmetric
counterterms upon projection on the action-like sector. 
At orders higher than one in the loop expansion there are additional
contributions to the non-invariant counterterms, arising from
known lower order terms. They can also be parameterized by using
the ST functionals.
We apply the method to  Yang-Mills theory 
in the Landau gauge with an explicit mass term introduced
in a BRST-invariant way via a BRST doublet. Despite being non-unitary,
this model provides a good example where the method devised in the paper
can be applied to derive the most general solution for the action-like part
of the quantum effective action, compatible with the fulfillment
of the ST identities, the ghost equation, the anti-ghost equation
and the $B$-equation, to all orders in the loop expansion.
We also provide the full dependence of the solution on the normalization conditions, that have to be supplemented in order to fix
the parameters left free in the symmetric quantum effective action
after the imposition of the ST identities
and of the other relevant symmetries of the model.
\end{quotation}

\newpage

\section{Introduction}

In the BRST quantization of gauge theories the BRST invariance
of the gauge-fixed classical action is translated at the quantum level
into the fulfillment of the Slavnov-Taylor (ST) identities
for the quantum effective action 
\cite{Becchi:1974md,Becchi:1974xu,Becchi:1975nq, Becchi:bd,
Piguet:er}.
As is well-known, the ST identities cannot be restored in the
case of anomalous gauge theories \cite{Piguet:er}. 
That is, there does not
exist any choice of finite non-symmetric counterterms, allowing to 
restore the ST identities at the quantum level.
As a result, the corresponding quantum theories fail to comply
with unitarity requirements 
\cite{Becchi:1974xu,Becchi:1975nq, Becchi:bd,Curci:1976yb, Kugo:zq}.

Even for non-anomalous theories
the regularization procedure, required to deal with UV divergences,
may break the ST identities at the quantum level. 
Unlike those of anomalous models, these breakings
are spurious, since they can be removed by a suitable choice
of finite non-symmetric counterterms, order by order in the loop
expansion. The existence of the non-symmetric counterterms,
allowing to restore the ST identities at the quantum level,
and the complete algebraic characterization of the action-like
symmetric counterterms 
have been analyzed in a regularization-independent
way within the framework of Algebraic Renormalization
(for a review see \cite{Piguet:er}).
Along these lines 
pure Yang-Mills theory has been discussed for general gauge-fixing
condition in \cite{Kraus:1989xe}. 
Yang-Mills in the background field formulation is considered
in \cite{Grassi:1995wr}.
The complete renormalization structure 
of the Standard Model has been studied in
\cite{Kraus:1997bi,Grassi:1999nb}. Finally, the Minimal Supersymmetric Standard Model has been analyzed in \cite{Hollik:2002mv}.

In the Standard Model and in the Minimal Supersymmetric Standard Model
no regularization scheme is known to preserve {\em ab initio} and to
all orders in the loop expansion the ST identities, due to the presence
of the $\gamma_5$ matrix and of the completely antisymmetric tensor 
$\epsilon_{\mu\nu\rho\sigma}$.

We point out that the methods used in Algebraic Renormalization
in order to guarantee that the model is anomaly-free and to characterize 
by algebraic means the symmetric counterterms and the corresponding
normalization conditions \cite{Piguet:er} do not yield {\em per se} 
the non-symmetric counterterms, required in the explicit computations
of the Feynman amplitudes at the loop level.

The problem of the construction of the non-symmetric counterterms
was addressed in \cite{Ferrari:1998jy,Ferrari:1999nj}.  
Under the assumption that the ST identities have been recursively
restored up to order $n-1$ in the loop expansion, 
by the Quantum Action Principle
\cite{Lam:1972mb,Lam:qk,Lam:qa,Lowenstein:1971jk,Breitenlohner:hr}
the most general $n$-th order ST breaking functional $\Delta^{(n)}$ is a local
integrated  functional in the fields, the antifields and their derivatives with
bounded dimension.
$\Delta^{(n)}$ can be explicitly computed and  the non-symmetric
counterterms, designed to reabsorb $\Delta^{(n)}$, can be worked out.
This was done for the Abelian Higgs-Kibble model in \cite{Ferrari:1998jy}.

In \cite{Ferrari:1999nj}
it was shown that the computation of the $n$-th order breaking terms
can actually be avoided: for anomaly-free models and under the assumptions
that the ST identities have been recursively restored up to order $n-1$,
the direct imposition of the $n$-th order ST identities at the level
of the symmetric $n$-th order effective action is shown to be equivalent,
in the absence of IR problems,
to the solution of a set of linear equations, whose unknowns are 
the $n$-th order coefficients of the action-like monomials.
These coefficients turn
out to be functions of the $n$-th order superficially convergent 
Feynman amplitudes and known lower-order contributions.
The solution is non-unique: the most general solution is obtained by 
adding all possible symmetric action-like
counterterms to a particular solution of the set of linear equations.
The ambiguities must be fixed by supplementing suitable normalization
conditions \cite{Ferrari:1998jy,Ferrari:1999nj}.

This method has a very general range of applicability and does not rely on
the existence of a nilpotent classical linearized ST operator.
Indeed in \cite{Ferrari:1999nj} it was applied to the restoration of the
ST identities in the massive Abelian Higgs-Kibble model,
where the linearized classical ST operator is not nilpotent, due
to the presence of an explicit mass term for the Abelian gauge field.

In \cite{Steinhauser,Grassi:2001zz,Grassi:2000kp,Grassi:2001kw}
this technique was applied to the  
case of Standard Model processes. Supersymmetric
models have been analyzed in \cite{Hollik:2000pa,Hollik:1999xh}.

In this paper we will analyze an alternative algorithm for the computation 
of the non-symmetric counterterms, required in order to recursively 
restore the 
ST identities, which can be applied whenever the linearized classical
ST operator ${\cal S}_0$ is nilpotent.
This method provides a geometrical characterization of the non-invariant
counterterms: it is shown that the recursive
imposition of the ST identities can be equivalently 
formulated as a parameterization problem for the local
approximation of the $n$-th order symmetric quantum effective action 
$\GG^{(n)}$ in terms of 
ST functionals, associated with the cohomology classes of the classical
linearized ST operator ${\cal S}_0$.

This allows for an effective characterization of both the symmetric
and the non-symmetric
counterterms, needed to restore the ST identities at order $n\geq 1$ in the
loop expansion, as it will be explained in Sect.~\ref{STparam}.

It turns out that solving the parameterization problem is easier than directly solving 
the problem of the recursive restoration of the STI. In particular, the parameterization problem
is more suited to be implemented via symbolic computation \cite{prep}.

In this paper we focus on the geometrical aspects of this procedure.
We analyze the simple case of Yang-Mills theory 
in the Landau gauge
with an explicit mass term introduced in a BRST-invariant way
via a BRST doublet \cite{Blasi:1995vt}. Despite being non-unitary
at non-zero mass, this model provides a good example
where the method described in this paper can be applied to derive
the most general solution for the action-like part of the quantum
effective action, compatible with the fulfillment of the ST
identities, the ghost equation, the antighost equation and the $B$-equation,
to all orders in the loop expansion.
In particular, since all dimensionful parameters enter into
 the classical action 
via a BRST-exact term, many
simplifications arise in the solution of the parameterization problem.

The analysis of  those models where
dimensionful parameters are introduced in the classical action
via cohomologically non-trivial terms (like in spontaneously
broken gauge theories)  only involves additional computational complications
in the solution of the parameterization problem, which are better dealt 
with by computerized algorithms \cite{prep}. 
In particular, it turns out that the
construction of the action-like part of $\GG^{(n)}$ in terms of the
ST functionals remains comparatively simple \cite{prep}.

\medskip

The paper is organized as follows. In Sect.~\ref{STparam} we 
discuss how the recursive restoration of the ST identities
for anomaly-free gauge theories
can be translated into a parameterization problem for the 
local approximation to the symmetric quantum effective action
in terms of suitable ST functionals.
These ST functionals are associated with the cohomology classes of ${\cal S}_0$
 in the
space of local functionals without action-like power-counting restriction. 
This in turn provides
useful simplifications in the computation of the non-invariant counterterms, needed to restore 
the ST identities at higher order in the loop expansion. Moreover, 
this provides a geometrical insight
into the structure of the local part of the symmetric quantum effective action, relevant for the restoration
of the ST identities.
By using the method described in Sect.~\ref{STparam} the close connection between the ST breaking terms and
the existence of mass scales in the theory becomes apparent. 

In Sect.~\ref{YM} we discuss  Yang-Mills theory 
in the Landau gauge with an explicit mass term 
introduced in a BRST-invariant way
via a BRST doublet.
 We derive the most general form of the action-like part of the quantum
effective action, compatible with the fulfillment of the ST identities,
the ghost equation, the antighost equation and the $B$-equation, to all
order in the loop expansion, and discuss how to recursively construct
the non-symmetric $n$-th order counterterms, needed to restore 
the ST identities possibly broken by the regularized $n$-th order
quantum effective action.

Finally conclusions are presented in Sect.~\ref{conc}.

\section{ST parameterization of the quantum effective action}\label{STparam}

In this section we will describe an effective method for the 
restoration at the quantum level of the Slavnov-Taylor (ST) identities
\begin{eqnarray}
{\cal S}(\GG) \equiv (\GG,\GG) = \sum_i
		\int d^4x \, \frac{\delta \GG}{\delta \Phi_i^*}
		             \frac{\delta \GG}{\delta \Phi_i} = 0
\label{intro1}
\end{eqnarray}
for anomaly-free gauge theories. 

The symmetric quantum effective action
$\GG$ in eq.(\ref{intro1}) is a functional of the fields $\Phi_i$ and of the
corresponding antifields $\Phi_i^*$, coupled in the 
BRST-invariant classical action $\GG^{(0)}$
to the BRST variation of the fields $\Phi_i$. 
For the sake of definiteness, we will restrict ourselves to the case
of  closed gauge algebras.
The fields $\Phi_i$ include the gauge and matter fields, collectively
denoted by $\varphi_k$, the ghost
fields $\omega^\alpha$ and the fields $\bar \omega^\alpha,B^\alpha$, belonging
to the non-minimal sector needed to fix the gauge \cite{Gomis:1994he}. 
Their BRST transformations are 
\begin{eqnarray}
s \varphi_k = R^\alpha_k[\varphi]\omega_\alpha \, , ~~~~~
s \omega_\alpha = \frac{1}{2} F^{\beta\gamma}_\alpha \omega_\beta \omega_\gamma \, , 
~~~~~ s \bar \omega^\alpha = B^\alpha \, , ~~~~~ s B^\alpha = 0 \, .
\label{thg1}
\end{eqnarray}
The constants $F^{\beta\gamma}_\alpha$ are antisymmetric
in the indices $\beta,\gamma$ and are related to $R^\alpha_k$
by the algebra
\begin{eqnarray}
R^\alpha_j[\varphi] \frac{\delta R^\beta_k[\varphi]}{\delta \varphi_j}
- 
R^\beta_j[\varphi] \frac{\delta R^\alpha_k[\varphi]}{\delta \varphi_j}
= F^{\alpha\beta}_\gamma R^\gamma_k[\varphi] \, .
\label{thg2}
\end{eqnarray}
The Jacobi identity
\begin{eqnarray}
F^{\beta[\gamma}_\alpha F^{\delta\sigma]}_\beta=0 \, 
\label{thg3}
\end{eqnarray}
holds true for the structure constants $ F^{\alpha\beta}_\gamma$.
We assign mass dimension $+1$ to the fields $\omega^\alpha$ and
$\bar \omega^\alpha$ and mass dimension $+2$ to $B^\alpha$.

From the point of view
of renormalization, the antifields $\Phi_i^*$ are only needed for
those fields whose BRST variation is non-linear in the quantum fields
\cite{Zinn_Justin}.
However, in order to keep the notation uniform, we assume that
an antifield has been introduced for each field of the model.
This allows to write the ST identities in the form of eq.(\ref{intro1}).

The symmetric quantum effective action
$\GG$ admits an expansion in the loop parameter $\hbar$:
\begin{eqnarray}
\GG = \sum_{n=0}^\infty \GG^{(n)} \, . 
\label{intro2}
\end{eqnarray}
In the above equation
$\GG^{(n)}$ stands for the coefficient  of $\GG$ of order $n$ in the $\hbar$
expansion.

Let us assume that the ST identities in eq.(\ref{intro1}) have 
been fulfilled up to order $n-1$, so that
\begin{eqnarray}
{\cal S}(\GG)^{(j)} = 0 \, , ~~~~~~~~~ j=0,1,\dots,n-1 \, .
\label{intro7}
\end{eqnarray}
At the $n$-th order in the loop expansion the ST identities read
\begin{eqnarray}
{\cal S}_0(\GG^{(n)}) =- \sum_{k=1}^{n-1} (\GG^{(n-k)},\GG^{(k)}) \, ,
\label{intro8}
\end{eqnarray}
where ${\cal S}_0$ is the classical linearized ST operator
\begin{eqnarray}
{\cal S}_0 = \sum_i \int d^4x \, \left ( 
\frac{\delta \GG^{(0)}}{\delta \Phi_i^*} \frac{\delta}{\delta \Phi_i}
+
\frac{\delta \GG^{(0)}}{\delta \Phi_i} \frac{\delta}{\delta \Phi_i^*}
\right ) \, .
\label{intro9}
\end{eqnarray}
The bracket in the R.H.S. of eq.(\ref{intro8}) is given by
\begin{eqnarray}
(X,Y) = \sum_i \int d^4x \, \frac{\delta X}{\delta \Phi^*_i} 
                     \frac{\delta Y}{\delta \Phi_i} \, .
\label{intro10}
\end{eqnarray}

Whenever the linearized classical ST operator ${\cal S}_0$
is nilpotent,
we will  prove
that the problem of the fulfillment of anomaly-free ST identities
at the $n$-th order in the loop expansion
can be equivalently formulated as the problem of parameterizing
the local approximation to 
the quantum effective action in terms of ST functionals,
associated with the cohomology classes of the classical
linearized ST operator. 
Upon projection of $\GG^{(n)}$ on the action-like sector, 
we construct the full $n$-th order action-like part of the
symmetric quantum effective action, fulfilling the ST identities
up to order $n$. 

The ${\cal S}_0$-invariants with dimension $\leq 4$ provide 
the usual symmetric counterterms. Their coefficients are not fixed
by the ST identities and enter as free parameters into $\GG^{(n)}$. 
There may exist additional relations between these parameters, stemming
from additional symmetries obeyed by the quantum
effective action like for instance the ghost equation, discrete
symmetries like C-parity, etc.
Once these relations have been taken into account, 
the remaining free parameters
must be fixed by supplying a suitable set of normalization conditions 
\cite{Piguet:er}.

\medskip
We will first briefly review the construction given in 
\cite{Ferrari:1999nj}, since this helps
clarifying the novel features of the method we propose 
in this paper and its range of applicability.
We are only concerned here with the construction
of the action-like part of $\GG$;  the first non-vanishing order
in the loop expansion of the ST identities  is local,  due to the QAP.
Hence we can restrict ourselves to the space of local functionals
and consider the effective local approximation to $\GG$.
That is, we associate to $\GG^{(n)}$ a formal power series
given by an infinite sum of local Lorentz-invariant
functionals. We will denote the series by $\GG^{(n)}$
itself:
\begin{eqnarray}
\GG^{(n)} = \sum_j \int d^4x \, m_j^{(n)} {\cal M}_j(x) \, .
\label{e_series}
\end{eqnarray}
In the absence of IR problems, as it happens for the massive
Abelian Higgs-Kibble model analyzed in \cite{Ferrari:1999nj},
$\GG^{(n)}$ in eq.(\ref{e_series})
is obtained by performing the Taylor expansion of the symmetric 1-PI amplitudes
in the independent external momenta around zero.

The Lorentz-scalar monomials ${\cal M}_j(x)$ in the fields and external
sources (and their derivatives) have to comply with all unbroken symmetries
of the theory and must have ghost number zero.
${\cal M}_j(x)$ are chosen to be linearly independent. They span
the vector space ${\cal V}$, to which $\GG^{(n)}$ in eq.(\ref{e_series})
belongs.
Notice that we do not impose any power-counting restriction
in the R.H.S. of eq.(\ref{e_series}): the basis $\{ {\cal M}_j \}_{j \in {\bf N}}$
contains elements with arbitrary positive dimension. Once the basis
$\{ {\cal M}_j \}_{j \in {\bf N}}$ has been chosen,
the coefficients $m_j^{(n)}$ in eq.(\ref{e_series}) are uniquely
determined.
We denote by $\gamma_j^{(n)}$ those coefficients in eq.(\ref{e_series}),
associated to monomials ${\cal M}_j(x)$ with dimension strictly
greater than four. They correspond to superficially convergent 
Feynman amplitudes. We denote by $\xi^{(n)}_j$ the coefficients
of the monomials ${\cal M}_j$ with dimension less or equal than four
(action-like part of $\GG^{(n)}$).

We always assume that the ST identities have been restored up to order
$n-1$, i.e.
\begin{eqnarray}
{\cal S}(\GG)^{(j)} = 0 \, , ~~~~~ j= 0,1,\dots,n-1 \, .
\label{e2_bis}
\end{eqnarray}
The $n$-th order ST identities read:
\begin{eqnarray}
{\cal S}(\GG)^{(n)} = {\cal S}_0 (\GG^{(n)}) + \sum_{j=1}^{n-1}
(\GG^{(n-j)},\GG^{(j)}) = 0 \, .
\label{e1}
\end{eqnarray}
The brackets in eq.(\ref{e1}) only involve $\GG^{(i)}$ with $i<n$, which
are known and fulfill eq.(\ref{e2_bis}).
The unknown quantities are the action-like terms of $\GG^{(n)}$ (i.e.
the  monomials with dimension less or equal to $4$
with the correct symmetry properties), which we denote by $\Xi^{(n)}$.
The regularized action
$\G^{(n)}$ is constructed by using the counterterms $\Xi^{(j)}$, $j<n$
and it is finite. We can expand $\Xi^{(n)}$ on a basis of
integrated Lorentz-invariant monomials with dimension $\leq 4$,
as follows:
\begin{eqnarray}
\Xi^{(n)} = \sum_j \int d^4x \, \xi_j^{(n)} {\cal M}^j(x) \, .
\label{e2}
\end{eqnarray}
${\cal S}(\GG)^{(n)}$ is an element of the vector space ${\cal W}$
spanned by all possible linearly independent Lorentz-invariant
monomials in the fields and the external sources and their derivatives
with ghost number $+1$, with possibly additional symmetry properties
dictated by the relevant symmetries of the model.
We denote by $\{ {\cal N}_i(x) \}_{i=1,2,3,\dots}$ a basis for ${\cal W}$.
Then we can insert the decomposition in eq.(\ref{e_series}) into eq.(\ref{e1})
and get
\begin{eqnarray}
\!\!\!\! {\cal S}(\GG)^{(n)} & = & \sum_j m_j^{(n)} {\cal S}_0 (\int d^4x \, 
{\cal M}_j) \nonumber \\
& & + \sum_{i=1}^{n-1} \sum_{jj'} m_j^{(i)} m_{j'}^{(n-i)}
\Big ( \int d^4x {\cal M}_j(x), \int d^4x' {\cal M}_{j'}(x') \Big ) \, .
\label{e3}
\end{eqnarray}
There exist coefficients $a_r^j, b_{kr}^{jj'}$ (uniquely fixed
by the choice of ${\cal N}_i(x)$ and by the action of ${\cal S}_0$)
such that
\begin{eqnarray}
{\cal S}_0 (\int d^4x \, {\cal M}_j) = \int d^4x \sum_r
a_r^j {\cal N}_r(x)  \, ,
\label{e4}
\end{eqnarray}
\begin{eqnarray}
\Big ( \int d^4x {\cal M}_j(x), \int d^4x' {\cal M}_{j'}(x') \Big ) =
\int d^4x \, \sum_r b_r^{jj'} {\cal N}_r(x) \, .
\label{e5}
\end{eqnarray}
Then eq.(\ref{e3}) becomes
\begin{eqnarray}
\sum_j a_r^j m_j^{(n)} + \sum_{i=1}^{n-1}
\sum_{jj'} m_j^{(i)} m_{j'}^{(n-i)} b_r^{jj'} = 0 \, , ~~~~
r=0,1,2\dots
\label{e6}
\end{eqnarray}
For $r$ such that ${\rm dim} \, {\cal N}_r(x)>5$ 
(with the conventions on the dimensions given in the Introduction),
eq.(\ref{e6}) is an identity by the virtue of the Quantum
Action Principle (QAP). 
For $r$ such that ${\rm dim} \, {\cal N}_r \leq 5$
 eq.(\ref{e6}) defines an inhomogeneous linear problem
in the unknowns $\xi_j^{(n)}$.
In the absence of anomalies,
one can solve eq.(\ref{e6}) by expressing the coefficients
$\xi_j^{(n)}$ in terms of the coefficients $\gamma_j^{(n)}$ 
of the $n$-th order superficially convergent Feynman amplitudes
and of the lower-order coefficients $m_j^{(l)}$, $l<n$. 
That is, one can construct
$\Xi^{(n)}$ from the superficially convergent part of $\GG^{(n)}$
and from lower order contributions $\GG^{(l)}$, $l<n$.

\par
The solution is not unique, due to the fact that the kernel
of ${\cal S}_0$ is not empty. The ambiguities are parameterized
by the action-like symmetric counterterms, possibly restricted
by further symmetries of the quantum effective action like
C-parity, the ghost equation, and so on. These ambiguities must be fixed
by providing a suitable set of normalization conditions.
\par

This procedure has the advantage 
that it can be applied to very general situations.
Indeed one can also drop the requirement of nilpotency of ${\cal S}_0$
and still the linear problem in eq.(\ref{e6}) remains well-defined.
We remark that in \cite{Ferrari:1999nj} this method was applied to the massive Abelian
Higgs-Kibble model, whose associated classical linearized
ST operator ${\cal S}_0$ is not nilpotent, due to the explicit mass term
for the Abelian gauge field.

\medskip

In those cases where ${\cal S}_0$ is a truly nilpotent
differential operator, it turns out that it is possible
to write directly the solution to eq.(\ref{e6}) by exploiting
the corresponding homogeneous equation, to be solved in a space
of local functionals not restricted by power-counting.
Since the latter defines a cohomological problem in the space of local functionals,
the powerful techniques of homological perturbation theory 
\cite{henn1,Barnich:mt,Barnich:db,Barnich:ve,Henneaux:ig} 
can be
used to obtain the required solution.
The action-like part of $\GG^{(n)}$ is then recovered
by projecting the solution onto the action-like sector.

In what follows $\G^{(n)}$ stands 
for the $n$-th order regularized vertex functional.
We first discuss the one-loop order. Eq.(\ref{e1}) reads at the one-loop
level:
\begin{eqnarray}
{\cal S}_0 (\GG^{(1)}) = 0 \, .
\label{n1}
\end{eqnarray}
$\GG^{(1)}$ in eq.(\ref{n1}) denotes the general solution
fulfilling the one-loop ST identities. We do not impose
for the moment any normalization condition. Hence 
$\GG^{(1)}$ will depend on a set of free parameters,
in one-to-one correspondence with the independent 
symmetric counterterms of the model at hand.

In the absence of anomalies, we know that for any regularized
effective action $\G^{(1)}$ there exists a local
action-like counterterm functional ${\Upsilon}^{(1)}$ such that
\begin{eqnarray}
\GG^{(1)} = \G^{(1)} + {\Upsilon}^{(1)} \, .
\label{n2}
\end{eqnarray}

Since ${\Upsilon}^{(1)}$ is action-like, $(1-t^4)\Upsilon^{(1)} = 0$.
$t^4 X$ denotes the projector on the action-like sector of the functional
$X$.
From eq.(\ref{n2}) we get that
\begin{eqnarray}
(1 - t^4)\G^{(1)} = (1-t^4) \GG^{(1)} \, .
\label{n5}
\end{eqnarray}
That is, the non action-like part of $\GG^{(1)}$ is fixed by the regularized
one-loop effective action.
The direct construction of the action-like part of $\GG^{(1)}$
proceeds as follows. 
Since we are only concerned with the construction
of $t^4 \GG^{(1)}$ and the first non-vanishing order
of the ST identities in the loop expansion is local by the QAP, 
we can restrict ourselves to the local approximation
given by eq.(\ref{e_series}).
Then we want to solve
\begin{eqnarray}
{\cal S}_0 (\GG^{(1)}) = 
{\cal S}_0 (t^4 \GG^{(1)} + (1 -t^4) \GG^{(1)}) = 0 
\label{inse6}
\end{eqnarray}
in the unknown $t^4 \GG^{(1)}$ (action-like part of $\GG^{(1)}$).
By eq.(\ref{inse6}) we see that the difference between $t^4 \GG^{(1)}$
and $-(1 -t^4) \GG^{(1)}$ must belong to the kernel of ${\cal S}_0$.
Since we are dealing with local functionals, a full characterization
of the kernel of ${\cal S}_0$ is available by means of cohomological
techniques \cite{henn1,Barnich:mt,Barnich:db,Barnich:ve,Henneaux:ig}. 
We stress that these techniques
are  independent of the power-counting restrictions. This plays an essential 
r\^ole here.

Any element ${\cal K}$ in the kernel of ${\cal S}_0$ 
in the sector with ghost number zero can be written as 
\begin{eqnarray}
{\cal K} = \sum_j \lambda_j^{(1)} {\Lambda}_j +
 {\cal S}_0({\cal K}^{(1)}_{\bf -1}) \, .
\label{inse7}
\end{eqnarray}
where the sum over $j$ is a sum over the cohomology classes 
${\Lambda}_j$ of ${\cal S}_0$ (not restricted by
power-counting), $\lambda^{(1)}_j$ are c-number
coefficients
and ${\cal K}^{(1)}_{\bf -1}$ is a local functional with ghost number $-1$.
We remark that ${\Lambda}_j$ are explicitly known 
for many gauge theories 
\cite{henn1,Barnich:mt,Barnich:db,Barnich:ve}. Recently 
this problem has also been solved for a wide class of 
four-dimensional supersymmetric models \cite{Brandt:2002pa}
.
From eq.(\ref{inse6}) and eq.(\ref{inse7}) we conclude that
\begin{eqnarray}
t^4 \GG^{(1)} = -(1-t^4)\GG^{(1)} + \sum_j \lambda_j^{(1)} {\Lambda}_j + {\cal S}_0({\cal K}^{(1)}_{\bf -1}) \, .
\label{inse8}
\end{eqnarray}
Now we apply the projector $t^4$ to both sides of eq.(\ref{inse8}) 
(this takes into account the dimension bounds on the functionals) and we
get
\begin{eqnarray}
\Xi^{(1)} & = & t^4 \GG^{(1)} = \sum_j \lambda_j^{(1)} t^4 {\Lambda}_j
+ t^4 \left ({\cal S}_0({\cal K}_{\bf -1}) \right ) \nonumber \\
          & = & \sum_j \lambda_j^{(1)} t^4 {\Lambda}_j
+ t^4 \left ( {\cal S}_0 (\sum_r \rho^{(1)}_r {\cal K}_{{\bf -1},r} \right )
\, .
\label{inse9}
\end{eqnarray}
In the second line of the above equation we have expanded the 
functional ${\cal K}^{(1)}_{\bf -1}$ on a basis $\{ {\cal K}_{{\bf -1},r} \}$
for the sector with ghost number $-1$.

The formula in eq.(\ref{inse9}) 
provides an explicit solution for $t^4 \GG^{(1)}$ 
in terms of local functionals belonging to the kernel of ${\cal S}_0$.
The coefficients $\lambda_j^{(1)}$, associated with
representatives ${\Lambda}_j$ whose dimension is $\leq 4$
\footnote{If $P = \sum_k {\cal Q}_k$ is a polynomial in the fields,
the external sources and their derivatives, decomposed into the sum
over the monomials ${\cal Q}_k$, we define the dimension of ${\cal P}$
as the maximum of the dimensions of ${\cal Q}_k$.},
are not constrained by the ST identities. They are free parameters
entering into the solution. They may not be all independent, due to the
existence of additional symmetries of the model (like the ghost equation
and discrete symmetries \cite{Piguet:er}). 
On the contrary, the coefficients $\lambda_j^{(1)}$ of those 
invariants whose dimension is greater than $4$ are not free: they can
be expressed in terms of the one-loop superficially  convergent Feynman
amplitudes entering into $(1-t^4)\G^{(1)} = (1-t^4)\GG^{(1)}$.
In the same way, the coefficients $\rho^{(1)}_r$, associated
to those ${\cal S}_0$-invariants ${\cal S}_0 (  {\cal K}_{{\bf -1},r} )$
whose dimension is $\leq 4$, are not constrained by the ST identities
and are free parameters entering into the solution.
Again there may exist relations between them, due to additional symmetries
of the model. The coefficients $\rho^{(1)}_r$, associated
to those ${\cal S}_0$-invariants ${\cal S}_0 (  {\cal K}_{{\bf -1},r} )$
with dimension $>4$, are fixed by the known non-action-like part of $\GG^{(1)}$
$(1-t^4)\GG^{(1)}=(1-t^4)\G^{(1)}$.
We notice that the coefficients $\xi^{(1)}_j$ in eq.(\ref{e2}) 
are in general functions of the coefficients $\lambda_j^{(1)}$
and $\rho^{(1)}_r$ in eq.(\ref{inse9}).
The non-symmetric counterterms, entering into $t^4\GG^{(1)}$,
 are given by
\begin{eqnarray}
\Xi^{(1)}_{n.s.} = {\sum_j}^\prime \lambda_j^{(1)} t^4 {\Lambda}_j
+ t^4 \left ( {\cal S}_0 ({\sum_r}^\prime \rho^{(1)}_r {\cal K}_{{\bf -1},r}) \right )
\, ,
\label{nonsimm1}
\end{eqnarray}
where the primed sum over $j$ is restricted to those invariants
$\Lambda_j$ with dimension $>4$ and the primed sum over $r$
is restricted to those invariants ${\cal S}_0({\cal K}_{{\bf -1},r})$
with dimension $>4$.

If the classical action contains dimensionful parameters like
masses it can happen \cite{prep} that
$\Xi^{(1)}_{n.s.}$ is non-zero.
As an example, this accounts for the dependence of the 
non-invariant counterterms on the Higgs v.e.v. $v$ 
in the case of the Abelian
Higgs-Kibble model analyzed in \cite{Ferrari:1998jy}.

The method of the parameterization 
in terms of ST functionals allows to derive easily the
structure of the symmetric solution $\GG^{(1)}$. 
Moreover, it 
provides a direct control of the dependence on the normalization
conditions chosen.

Let us now analyze how this procedure can be extended to higher
orders. The $n$-th order ST identities are
\begin{eqnarray}
{\cal S}_0 (\GG^{(n)}) + \sum_{k=1}^{n-1} ( \GG^{(k)}, \GG^{(n-k)} ) = 0 \, .
\label{h1}
\end{eqnarray}
Due to the fact that ${\cal S}_0^2=0$, by applying ${\cal S}_0$ to both
sides of eq.(\ref{h1}) we get
\begin{eqnarray}
{\cal S}_0 \Big ( \sum_{k=1}^{n-1} ( \GG^{(k)}, \GG^{(n-k)} ) \Big ) = 0 \, .
\label{h2}
\end{eqnarray}
By virtue of eq.(\ref{h2}) we get that
\begin{eqnarray}
\sum_{k=1}^{n-1} ( \GG^{(k)}, \GG^{(n-k)} ) =
\sum_{j \in {\cal J}} \lambda_j^{(n)} {\cal A}_j + 
{\cal S}_0( \tilde \zeta_{\bf 0}^{(n)} ) \, .
\label{h3}
\end{eqnarray}
In the above equation the sum over $j \in {\cal J}$ is a sum over
the representatives of the non-trivial cohomology classes
of ${\cal S}_0$ with ghost number $+1$ (anomalies)
and $\tilde \zeta_{\bf 0}^{(n)}$ is
a functional with ghost number zero.

Let us assume that no anomalies are present due to purely algebraic
reasons, so that the set ${\cal J}$ is empty. Hence eq.(\ref{h3}) reduces
to 
\begin{eqnarray}
\sum_{k=1}^{n-1} ( \GG^{(k)}, \GG^{(n-k)} ) = 
 {\cal S}_0( \tilde \zeta_{\bf 0}^{(n)} ) 
\label{h4}
\end{eqnarray}
and eq.(\ref{h1}) now becomes
\begin{eqnarray}
{\cal S}_0 ( \GG^{(n)} + \tilde \zeta^{(n)}_{\bf 0} ) = 0 \, ,
\label{h5}
\end{eqnarray}
which parallels eq.(\ref{n1}). Notice that $\tilde \zeta^{(n)}_{\bf 0}$ is
a known functional constructed from $\GG^{(i)}$, $i<n$, which are known.
By following the same approach described before for the one-loop
level we get from eq.(\ref{h5})
\begin{eqnarray}
{\cal S}_0[ t^4 ( \GG^{(n)} + \tilde \zeta^{(n)}_{\bf 0} )+
(1 -t^4) ( \GG^{(n)} + \tilde \zeta^{(n)}_{\bf 0} ) ] = 0 
\label{h6}
\end{eqnarray}
or 
\begin{eqnarray}
 t^4 ( \GG^{(n)} + \tilde \zeta^{(n)}_{\bf 0} ) & = & 
-(1 -t^4) ( \GG^{(n)} + \tilde \zeta^{(n)}_{\bf 0} ) \nonumber \\
&& + \sum_j \lambda_j^{(n)} {\Lambda}_j + {\cal S}_0({\cal K}_{\bf -1}^{(n)}) \, .
\label{h7}
\end{eqnarray}
By applying the projection operator $t^4$ to eq.(\ref{h7}) we get
\begin{eqnarray}
t^4 \GG^{(n)} = -t^4 \tilde \zeta^{(n)}_{\bf 0} + 
\sum_j \lambda_j^{(n)} t^4 {\Lambda}_j + t^4 {\cal S}_0(
\sum_r \rho_r^{(n)} {\cal K}_{{\bf -1},r} ) \, ,
\label{h8}
\end{eqnarray}
which gives explicitly the action-like part of $\GG^{(n)}$ 
to orders $n>1$.
The $n$-th order non-symmetric counterterms entering into $t^4\GG^{(1)}$
are given by
\begin{eqnarray}
\Xi^{(n)}_{n.s.} = - t^4 \tilde \zeta_{\bf 0}^{(n)} + {\sum_j}^\prime
\lambda_j^{(n)} t^4 \Lambda_j + t^4 {\cal S}_0
\left ( {\sum_r}^\prime \rho_r^{(n)} {\cal K}_{{\bf -1},r} \right )
\label{nonsymmn}
\end{eqnarray}
where the primed sum over $j$ is restricted to those invariants
$\Lambda_j$ with dimension $>4$ and the primed sum over
$r$ is restricted to those invariants ${\cal S}_0({\cal K}_{{\bf -1},r})$
with dimension $>4$.
Notice the appearance of non-symmetric counterterms depending on lower
orders via the term $-t^4 \tilde \zeta_{\bf 0}^{(n)}$ in the R.H.S.
of eq.(\ref{nonsymmn}).

Let us comment on the results of this subsection. 
The problem of the restoration of the
ST identities has been turned into a parameterization problem for 
the local approximation to the symmetric quantum effective action. 
It can be parameterized by means of the ST functionals $\Lambda_j$ and
${\cal S}_0( {\cal K}_{{\bf -1},r})$.
The lower order contributions to the ST identities given by eq.(\ref{h4})
are parameterized via the functional $\tilde \zeta^{(n)}_{\bf 0}$. 

Then the full action-like part of the $n$-th order symmetric
quantum effective action $t^4 \GG^{(n)}$ is provided by eq.(\ref{h8}).

\section{Yang-Mills theory with explicit BRST-invariant mass term}\label{YM}

In this section we apply the method described in Sect.~\ref{STparam}
to Yang-Mills theory 
in the Landau gauge 
with an explicit mass term introduced
in a BRST-invariant way via a BRST doublet \cite{Blasi:1995vt}.
At mass different than zero this model is not unitary. 
Nevertheless, it provides a good example to illustrate
the method analyzed in Sect.~\ref{STparam}. 
This is due to the fact that,
since all dimensionful parameters enter into the classical action
via a BRST-exact term, 
some simplifications arise
in the structure of the non-symmetric counterterms 
in $t^4\GG^{(n)}$ in eq.(\ref{nonsimm1})
and in eq.(\ref{nonsymmn}).
In particular,  there are no contributions 
from the $\Lambda$-type invariants
to the non-symmetric counter-terms in eqs.(\ref{nonsimm1})
and (\ref{nonsymmn}). 
However there are explicit
contributions to the non-symmetric counterterms 
in $t^4\GG^{(n)}$ at orders higher than one,
coming from $-t^4 \tilde \zeta^{(n)}_{\bf 0}$ in eq.(\ref{nonsymmn}).

We will show how the parameterization problem for the construction
of $\tilde \zeta^{(n)}_{\bf 0}$ can be solved by using techniques of homological
perturbation theory and how the method described in Sect.~\ref{STparam}
can be used to control the full dependence of $t^4 \GG^{(n)}$ 
on the normalization conditions chosen.

We do not discuss here the zero-mass limit. 

\medskip

We start from the classical action of Yang-Mills theory based on the 
simple compact gauge group $G$:
\begin{eqnarray}
S_{\rm YM} & = & \int d^4x \, \left \{ -\frac{1}{4g^2} G_{\mu\nu}^a G^{\mu\nu \, a}
- \bar \omega^a \partial_\mu
(D^\mu \omega)^a + B^a \partial A^a \right . \nonumber \\
&& \left . ~~~~~~~~~~ + A_\mu^{a*} (D^\mu \omega)^a - \omega^{* a} \frac{1}{2}f^{abc}
     \omega^b \omega^c + 
     \bar \omega^{a*} B^a 
   \right \} \, .
\label{ym1}
\end{eqnarray}
The non-Abelian gauge field strength $G_{\mu\nu}^a$ is
\begin{eqnarray}
G_{\mu\nu}^a = \partial_\mu A_\nu^a - \partial_\nu A_\mu^a
+ f^{abc} A_\mu^b A_\nu^c \, ,
\label{ym2}
\end{eqnarray}
while the covariant derivative $D^\mu_{ac}$ in the adjoint representation 
is given by 
\begin{eqnarray}
D^\mu_{ac} = \delta_{ac} \partial^\mu + f_{abc} A^\mu_b \, .
\label{ym3}
\end{eqnarray}
$f^{abc}$ are  the structure constants of the gauge group, $g$ is the coupling
constant of the model.
$\omega^a$ are the ghost fields, $\bar \omega^a$ the antighost fields,
$B^a$ the Nakanishi-Lautrup multiplier fields. 

The BRST differential $s$ acts as follows on the fields of the model:
\begin{eqnarray}
&& s A_\mu^a = (D_\mu \omega)^a \, , ~~~~ 
s \omega^a = -\frac{1}{2} f^{abc} \omega^b \omega^c \, , ~~~~ \nonumber \\
&& s \bar \omega^a = B^a \, , ~~~~ s B^a = 0 \, .
\label{ym8}
\end{eqnarray}
In $S_{\rm YM}$ we have coupled the BRST variations of $A_\mu^a,\omega^a,
\bar \omega^a$
to the antifields $A_\mu^{a*}, \omega^{a*}, \bar \omega^{a*}$.
The dimension of $A_\mu^a,\omega^a,\bar\omega^a$ is one,
the dimension of $B^a,A_\mu^{a *},\omega^{a*}, \bar \omega^{a*}$ is two.

The ghost number is assigned as follows:
\begin{eqnarray}
&& gh(A_\mu^a)=gh(B^a)
=gh(\bar \omega^{a*}) = 0 \, , ~~~~~
gh(\omega^a)=+1 \, ,  \nonumber \\
&& gh(\bar \omega^a)=gh(A^{a*}_\mu)
=-1 \, ,
~~~~~ gh(\omega^{a*})= -2 \, .
\label{ym8_bis} 
\end{eqnarray}
We now introduce an anticommuting parameter $\bar \rho$ with dimension
zero and ghost number $-1$ and its corresponding BRST partner
$m$ with dimension $+1$ and ghost number zero:
\begin{eqnarray}
s \bar \rho = m \, , ~~~~ s m = 0 \, .
\label{ym8_ter}
\end{eqnarray}
We consider the following classical action
\begin{eqnarray}
\G^{(0)} & = & S_{\rm YM} + \int d^4x \, s \left (
\frac{1}{2}\bar \rho m (A_\mu^a)^2 + \bar \rho m \bar \omega^a \omega^a 
\right ) \nonumber \\
& = & S_{\rm YM} +
\int d^4x \, \left ( \frac{1}{2} m^2 (A_\mu^a)^2 + m^2 \bar \omega^a \omega^a
\right . \nonumber \\
&   & \left . - \bar \rho m A_\mu^a \partial^\mu \omega^a - \bar \rho m B^a \omega^a - \frac{1}{2} \bar \rho m \bar \omega^a f^{abc} \omega^b \omega^c \right ) \, .
\label{ym8_quater}
\end{eqnarray}
Notice the appearance of a mass term for the gauge bosons
and  the ghost fields via the $s$-exact term in the first line of
eq.(\ref{ym8_quater}).
$\G^{(0)}$ is BRST-invariant:
\begin{eqnarray}
s \G^{(0)} = 0 \, .
\label{ym9}
\end{eqnarray}
As a consequence, $\G^{(0)}$ fulfills the ST identities:
\begin{eqnarray}
{\cal S}(\G^{(0)}) & = & \int d^4x \, \left (
\frac{\delta \G^{(0)}}{\delta A_\mu^{a*}} \frac{\delta \G^{(0)}}{\delta A_\mu^{a}}
+
\frac{\delta \G^{(0)}}{\delta \omega^{a*}} \frac{\delta \G^{(0)}}{\delta \omega^{a}} 
 + \frac{\delta \G^{(0)}}{\delta \bar \omega^{a*}}
   \frac{\delta \G^{(0)}}{\delta \bar \omega^a} \right ) 
+ m \frac{\partial \G^{(0)}}{\partial \bar \rho}
= 0 \, .
\label{ym10}
\end{eqnarray}
The linearized classical ST operator ${\cal S}_0$ is
\begin{eqnarray}
{\cal S}_0 & = & \int d^4x \, \left ( 
\frac{\delta \G^{(0)}}{\delta A_\mu^{a*}} \frac{\delta}{\delta A_\mu^{a}}
+
\frac{\delta \G^{(0)}}{\delta A_\mu^{a}} \frac{\delta}{\delta A_\mu^{a*}}
+
\frac{\delta \G^{(0)}}{\delta \omega^{a*}} \frac{\delta}{\delta \omega^{a}} 
+
\frac{\delta \G^{(0)}}{\delta \omega^{a}} \frac{\delta}{\delta \omega^{a*}} 
\right . \nonumber \\ 
&& 
\left . ~~~~~~~~~~~ + \frac{\delta \G^{(0)}}{\delta \bar \omega^{a*}}
                    \frac{\delta}{\delta \bar \omega^a} +
		    \frac{\delta \G^{(0)}}{\delta \bar \omega^a}
		    \frac{\delta}{\delta \bar \omega^{a*}}
\right ) 
+ m \frac{\partial}{\partial \bar \rho}
\, .
\label{ym11}
\end{eqnarray}
${\cal S}_0$ is nilpotent: ${\cal S}_0^2=0$.
We remark that the full dependence of $\G^{(0)}$ on $\bar \rho$
can be reabsorbed by performing the following antifield
redefinitions:
\begin{eqnarray}
&&  A^{a*'}_\mu = A^{a*}_\mu - \bar \rho m A^a_\mu \, , 
\nonumber \\
&& \omega^{a*'} = \omega^{a*} + \bar \rho m \bar \omega^a \, , 
\nonumber \\
&& \bar \omega^{a*'} = \bar \omega^{a*} - \bar \rho m \omega^a 
\, . 
\label{new1}
\end{eqnarray}
In the new variables $\G^{(0)}$ becomes
\begin{eqnarray}
\G^{(0)} 
& = & \int d^4x \, \left \{ -\frac{1}{4g^2} G_{\mu\nu}^a G^{\mu\nu \, a}
- \bar \omega^a \partial_\mu
(D^\mu \omega)^a + B^a \partial A^a \right . \nonumber \\
&& \left . + A_\mu^{a*'} (D^\mu \omega)^a - \omega^{* a'} \frac{1}{2}f^{abc}
     \omega^b \omega^c + 
     \bar \omega^{a*'} B^a 
     \right . \nonumber \\
&& \left . 
   + \frac{1}{2}m^2 (A_\mu^a)^2 + m^2 \bar \omega^a \omega^a 
   \right \} \, .
\label{new2}
\end{eqnarray}

$\G^{(0)}$ in eq.(\ref{new2}) obeys a set of additional identities:
\begin{eqnarray}
\frac{\partial \G^{(0)}}{\partial \bar \rho} = 0 \, ,
\label{new3}
\end{eqnarray}
the $B$-equation
\begin{eqnarray}
\frac{\delta \G^{(0)}}{\delta B^a} = \partial A^a 
+  \bar \omega^{a*'}
\, , 
\label{ym12}
\end{eqnarray}
the ghost equation
\begin{eqnarray}
\frac{\delta \G^{(0)}}{\delta \bar \omega^a} + 
\partial^\mu \frac{\delta \G^{(0)}}{\delta A^{a*'}_\mu} = m^2 \omega^a \, , 
\label{ym13}
\end{eqnarray}
and the anti-ghost equation
\begin{eqnarray}
&& \int d^4x \, \left ( \frac{\delta \G^{(0)}}{\delta \omega^a}
-f^{abc} \bar \omega^b \frac{\delta \G^{(0)}}{\delta B^c} \right )
\nonumber \\
&& ~~~~~ = \int d^4x \, \left ( m^2 \bar \omega^a 
- f^{abc} A_\mu^{b*'} A^{\mu c}
+ f^{abc} \omega^{*b'} \omega^c 
\right ) \, .
\label{ym14}
\end{eqnarray}
The $B$-equation and the ghost equation are local,
while the anti-ghost equation is an integrated equation.
We notice that the breaking terms in the R.H.S. of eqs.(\ref{ym13})
and (\ref{ym14}) are linear in the quantum fields.

All these functional identities can be restored at the quantum
level by applying the standard methods dicussed e.g. in \cite{Piguet:er}. 
Therefore we assume from now on that both the regularized
quantum effective action $\G$ and the symmetric quantum effective
action $\GG$ fulfill the following relations:
\begin{eqnarray}
\frac{\partial \Sigma^{(j)}}{\partial \bar \rho} = 0 \, , ~~~~~~~~~ j \geq 1 \, ,
\label{new4}
\end{eqnarray}
\begin{eqnarray}
\frac{\delta \Sigma^{(j)}}{\delta B^a} = 0 \, , ~~~~~~~~~ j \geq 1 \, ,
\label{ym15}
\end{eqnarray}
\begin{eqnarray}
\frac{\delta \Sigma^{(j)}}{\delta \bar \omega^a} 
+ \partial^\mu \frac{\delta \Sigma^{(j)}}{\delta A^{a*'}_\mu} = 0 \, ,
~~~~~~~~~ j \geq 1 \, ,
\label{ym16}
\end{eqnarray}
\begin{eqnarray}
\int d^4x \, \left ( \frac{\delta \Sigma^{(j)}}{\delta \omega^a}
-f^{abc} \bar \omega^b \frac{\delta \Sigma^{(j)}}{\delta B^c}
\right )  =
\int d^4x \, \frac{\delta \Sigma^{(j)}}{\delta \omega^a} = 0 \, ,
~~~~~~~~~ j \geq 1 \, .
\label{ym17}
\end{eqnarray}
In the above equations $\Sigma$ stands both for $\Gamma$ and 
$\GG$. In eq.(\ref{ym17})
we have used eq.(\ref{ym15}). A formal proof of the validity
of the identities (\ref{new4})-(\ref{ym17}) for $\GG$ can be achieved
by showing that there are no anomalies if the vertex functional
is restricted by eqs.(\ref{new4})-(\ref{ym17}). This can be done
by applying the standard methods described
for instance in \cite{Piguet:er}. Let us remark that, since
$\bar \rho$ and $m$ enter as a BRST doublet, they cannot modify
the cohomological properties of the model at $\bar \rho=m=0$.

From eq.(\ref{new4}) we see that $\G^{(j)}$ and $\GG^{(j)}$,
$j \geq 1$ are independent of $\bar \rho$.
From eq.(\ref{ym15}) we conclude that $\G^{(j)}$ and $\GG^{(j)}$, 
$j \geq 1$, are independent of $B^a$. From eq.(\ref{ym16}) 
we see that $\G^{(j)}$ and $\GG^{(j)}$, $j \geq 1$ depends on
$\bar \omega^a$ through the combination
\begin{eqnarray}
\hat A^{*a'}_\mu = A^{a*'}_\mu + \partial_\mu \bar \omega^a \, .
\label{ym18}
\end{eqnarray}

\subsection{One-loop level}\label{oneloop}

By following the method outlined in Sect.~\ref{STparam} 
we construct the most general
solution to the action-like part of the symmetric quantum effective action
$\GG^{(1)}$ at one-loop level, fulfilling the first order ST identities:
\begin{eqnarray}
{\cal S}_0 (\GG^{(1)}) = 0 \, .
\label{ym19}
\end{eqnarray}
Since $\GG^{(1)}$ is independent of $\bar \rho$ by virtue of eq.(\ref{new4}),
we only need to consider the restriction of ${\cal S}_0$ to the space
of $\bar \rho$-independent functionals.
By explicit computation one can verify that
\begin{eqnarray}
{\cal S}_0(\Phi^{*'}) = \frac{\delta \G^{(0)}_{m=0}}{\delta \Phi}\, ,
\label{add2_1}
\end{eqnarray}
for $\Phi^{*'} = \hat A^{a*'}_\mu, \omega^{a*'}, \bar \omega^{a*'}$.
In the primed variables 
the operator ${\cal S}_0$ has definite degree $+1$ with respect to the 
grading induced by the dimension of the fields and the antifields.
The assignment of dimensions to the primed variables is as follows:
$\hat A^{a*'}_\mu$, $\omega^{a*'}$, $\bar \omega^{a*'}$ have dimension $2$,
i.e. the dimension of the primed antifield is the same as the one of
the corresponding unprimed antifield.

Eq.(\ref{inse9}) now gives
the complete structure of the action-like part of $\GG^{(1)}$ in terms of the
invariants $\Lambda_j$ and ${\cal S}_0({\cal K}_{{\bf -1},r})$. 
The classification of the invariants $\Lambda_j$, relevant for the present
model, can be obtained according to the results of 
\cite{Piguet:er,henn1,Barnich:mt,Barnich:db,Barnich:ve}. In particular,
since there are no Abelian factors, no cohomologically non-trivial
invariants, involving the external sources 
$\hat A^{\mu *'}_a,\omega^{a*'},\bar \omega^{a*'}$
exist.

Due to the fact that ${\cal S}_0$ has degree $+1$, there are no
contributions to $t^4 \GG^{(1)}$ from ${\cal S}_0$-invariants
with dimension $\geq 5$. Hence at one loop level $t^4 \GG^{(1)}$ 
can be parameterized in terms of the following ${\cal S}_0$-invariants
with dimension $\leq 4$, whose coefficients are not fixed
by the ST identities:
\begin{eqnarray}
t^4 \GG^{(1)}  =  \lambda_1^{(1)} \int d^4x \, G_{\mu\nu}^a G^{\mu\nu}_a 
+ \rho_1^{(1)} {\cal S}_0 (\int d^4x \, 
\hat A^{a*'}_\mu A^a_\mu ) 
   + \rho_2^{(1)} {\cal S}_0 (\int d^4x \, 
\omega^{a*'} \omega^a ) 
\, . 
\label{ym20}
\end{eqnarray}
$t^4 \GG^{(1)}$ in eq.(\ref{ym20}) fulfills the $B$-equation and the ghost
equation. 
The anti-ghost equation in eq.(\ref{ym17}) fixes
the coefficient $\rho_2^{(1)}$ to zero. 
It does not constrain $\lambda_1^{(1)},\rho_1^{(1)}$.
These are free parameters entering into the most general solution
for $t^4 \GG^{(1)}$, compatible with the $B$-equation, the
ghost equation, the anti-ghost equation and the one-loop order
ST identities.
The parameters $\lambda_1^{(1)},\rho_1^{(1)}$
must be fixed
by supplying a set of normalization conditions for $t^4 \GG^{(1)}$. 
As an example, one might choose
\begin{eqnarray}
\xi^{(1)}_{G_{\mu\nu}^a G^{\mu\nu}_a} = 0 \, , ~~~~
\xi^{(1)}_{\hat A^{a*'}_\mu \partial^\mu \omega^a} = 0 \, .
\label{ym21}
\end{eqnarray}
yielding
\begin{eqnarray}
\lambda_1^{(1)} = 0 \, ,  ~~~~~ \rho_1^{(1)} = 0 \, .
\label{ym22}
\end{eqnarray}
In order to study the dependence of higher-order non-symmetric
counterterms on the normalization conditions, we do not restrict
ourselves to a particular set of normalization conditions. 
We will thus keep the explicit dependence on 
$\rho_1^{(1)}, \lambda_1^{(1)}$.
Finally we get
\begin{eqnarray}
t^4 \GG^{(1)} & = & \lambda_1^{(1)} \int d^4x \, G_{\mu\nu}^a G^{\mu\nu}_a 
+ \rho_1^{(1)} {\cal S}_0 (\int d^4x \, \hat A^{a*'}_\mu A^a_\mu ) 
\, .
 \label{ym23}
\end{eqnarray}
Only symmetric counterterms enter into $t^4 \GG^{(1)}$.
This is due to the fact that 
the only dimensionful parameter $m$ enters into $\G^{(0)}$ via a
$s$-exact term.
In theories where dimensionful parameters appear via
a cohomologically non-trivial term (like for instance in the
case of spontaneously broken models)
this property is not true \cite{prep}.
Also in Yang-Mills theory 
this feature is in general violated at higher
orders, as we will discuss in the next subsection. It can be
preserved only for a special choice of normalization conditions
for $t^4 \GG^{(1)}$.

\subsection{Higher loops}\label{higherorder}

At higher orders one also needs to take into account the contributions
from lower order terms, entering into $\tilde \zeta_{\bf 0}^{(n)}$ in
eq.(\ref{h4}). 
We can limit ourselves to the terms of dimension $\leq 5$ of 
the functional
\begin{eqnarray}
&& \Delta^{(n)} = \sum_{j=1}^{n-1} 
\left ( 
\frac{\delta \GG^{(n-j)}}{\delta \hat A^{a*'}_\mu} 
\frac{\delta \GG^{(j)}}{\delta A^a_\mu}
+
\frac{\delta \GG^{(n-j)}}{\delta \omega^{a*'}}
\frac{\delta \GG^{(j)}}{\delta \omega^a} 
+
\frac{\delta \GG^{(n-j)}}{\delta \bar \omega^{a*'}}
\frac{\delta \GG^{(j)}}{\delta \bar \omega^a}
\right ) \, .
\label{ym24}
\end{eqnarray}
$\Delta^{(n)}$ has ghost number $+1$. 
By taking into account the $B$-equation, the ghost equation
the anti-ghost equation and the fact that $\GG^{(j)}$ is independent
of $\bar \rho$, the part of  dimension $\leq 5$ of $\Delta^{(n)}$ 
(which from now on we will denote by $\Delta^{(n)}$ itself) reduces to
\begin{eqnarray}
\Delta^{(n)} & = & \sum_{j=1}^{n-1} \int d^4x \, 
\left ( \rho_1^{(n-j)}\rho_1^{(j)} (-\partial^\mu \omega^a)
\frac{\delta}{\delta A_\mu^a} {\cal S}_0 (\int d^4y \, 
\hat A^{b*'}_\nu A^{\nu b} ) 
\right . \nonumber \\
&& \left . ~~~~~~~~~~~~~
+ \rho_1^{(n-j)} \lambda_1^{(j)}  (-\partial^\mu \omega^a)
\frac{\delta}{\delta A_\mu^a} ( \int d^4y \, G^{\rho\sigma b}
G_{\rho\sigma}^b )
\right ) \nonumber \\
& = & - \int d^4x \, \partial^\mu \omega^a
\frac{\delta}{\delta A_\mu^a} 
\left [
\sum_{j=1}^{n-1} \rho_1^{(n-j)}\rho_1^{(j)}
{\cal S}_0 (\int d^4y \, \hat A^{b*'}_\nu A^{\nu b} ) 
\right . \nonumber \\
&  & ~~~~~~~~~~~ 
\left . + \sum_{j=1}^{n-1} \rho_1^{(n-j)} \lambda_1^{(j)}
 \int d^4y \, G^{\rho\sigma b} G_{\rho\sigma}^b 
\right ] \, .
\label{ym25}
\end{eqnarray}
The functional $\Delta^{(n)}$ fulfills the Wess-Zumino consistency
condition
\begin{eqnarray}
{\cal S}_0 (\Delta^{(n)}) = 0 \, .
\label{ym26}
\end{eqnarray}
In order to determine the non-symmetric higher order counterterms
we need to find a functional $\tilde \zeta^{(n)}_{\bf 0}$ such that
\begin{eqnarray}
{\cal S}_0 (\tilde \zeta^{(n)}_{\bf 0}) = \Delta^{(n)} \, .
\label{ym27}
\end{eqnarray}
We first consider the second 
term in the R.H.S. of eq.(\ref{ym25}). We set
\begin{eqnarray}
&& {\cal H}_2 = \sum_{j=1}^{n-1} \rho_1^{(n-j)} \lambda_1^{(j)}
 \int d^4y \, G^{\rho\sigma b} G_{\rho\sigma}^b \, .
\label{ym28}
\end{eqnarray}
${\cal H}_2$ 
does not depend on the antifields.
We first notice that the following formula holds for any functional
$H$:
\begin{eqnarray}
s \left ( \int d^4x \, \partial_\mu \omega_a 
\frac{\delta H}{\delta A_\mu^a} \right ) & = & 
\int d^4x \, \left ( - f^{abc} \partial_\mu \omega^b \omega^c 
\frac{\delta H}{\delta A_\mu^a} -
\partial_\mu \omega^a f^{abc} \omega^c 
\frac{\delta H}{\delta A_\mu^b} \right ) 
\nonumber \\
& & ~~~~~~~ - \int d^4x \, \partial_\mu \omega^a \frac{\delta}{\delta A_\mu^a}
(s H) \nonumber \\
& = & - \int d^4x \, \partial_\mu \omega^a \frac{\delta}{\delta A_\mu^a}
(s H)
\, .
\label{ym29}
\end{eqnarray}
Since $s {\cal H}_2=0$ 
we get that the second 
term in the R.H.S. of 
eq.(\ref{ym25}) separately satisfies the Wess-Zumino consistency condition
\cite{WZ,Piguet:er}
\begin{eqnarray}
{\cal S}_0 \int d^4x \, \left ( - \partial_\mu \omega_a
\frac{\delta}{\delta A_\mu^a} ( {\cal H}_2 ) \right ) = 
s \int d^4x \, \left ( - \partial_\mu \omega_a
\frac{\delta}{\delta A_\mu^a} ( {\cal H}_2 ) \right ) = 0 \, .
\label{ym30}
\end{eqnarray}
Moreover one gets that 
\begin{eqnarray}
s \left ( \int d^4x \, A^a_\mu \frac{\delta H}{\delta A_\mu^a} \right )
& = & \int d^4x \, \left ( \partial_\mu \omega_a 
\frac{\delta H}{\delta A_\mu^a} + f^{abc} A_\mu^b \omega^c 
\frac{\delta H}{\delta A_\mu^a} + A_\mu^a s \frac{\delta H}{\delta A_\mu^a}
\right ) \nonumber \\
& = & \int d^4x \, \left ( \partial_\mu \omega_a 
\frac{\delta H}{\delta A_\mu^a} + f^{abc} A_\mu^b \omega^c 
\frac{\delta H}{\delta A_\mu^a} + A_\mu^a [ s, \frac{\delta}{\delta A_\mu^a}] H + A_\mu^a \frac{\delta}{\delta A_\mu^a} (sH) 
\right ) \nonumber \\
& = & \int d^4x \, \left ( \partial_\mu \omega_a 
\frac{\delta H}{\delta A_\mu^a}
+ A^a_\mu \frac{\delta}{\delta A^a_\mu} (sH) \right ) \, .
\label{ym31}
\end{eqnarray}
Since $s {\cal H}_2=0$, the above formula allows to write
\begin{eqnarray}
\int d^4x \, \partial_\mu \omega_a \frac{\delta {\cal H}_2}{\delta A_\mu^a} 
= s \int d^4x \, A^a_\mu \frac{\delta}{\delta A^a_\mu} {\cal H}_2
= {\cal S}_0 \int d^4x \, A^a_\mu \frac{\delta}{\delta A^a_\mu} {\cal H}_2
\, .
\label{ym32}
\end{eqnarray}
This gives the terms in $\tilde \zeta^{(n)}_{\bf 0}$ generating
 the second term 
in the R.H.S. of eq.(\ref{ym25}):
\begin{eqnarray}
&& \!\!\!\!\!\!\!\!\!\!\! - \int d^4x \, \partial^\mu \omega^a
\frac{\delta}{\delta A_\mu^a} 
\left ( \sum_{j=1}^{n-1} \rho_1^{(n-j)} \lambda_1^{(j)}
 \int d^4y \, G^{\rho\sigma b} G_{\rho\sigma}^b 
\right ) \nonumber \\
&& = -
{\cal S}_0 (\int d^4x A_\mu^a \frac{\delta}{\delta A_\mu^a} 
\, \left [ 
\sum_{j=1}^{n-1} \rho_1^{(n-j)} \lambda_1^{(j)}  
\int d^4y \, G^{\rho\sigma b} G_{\rho\sigma}^b 
\right ] ) \, .
\nonumber \\ 
\label{ym33}
\end{eqnarray}
The analysis of the first term in the R.H.S. of eq.(\ref{ym25})
\begin{eqnarray}
{\cal H}_1 =  
\sum_{j=1}^{n-1} 
\rho_1^{(n-j)}\rho_1^{(j)} \int d^4x \, (-\partial^\mu \omega^a)
\frac{\delta}{\delta A_\mu^a} {\cal S}_0 (\int d^4y \, 
\hat A^{b*'}_\nu A^{\nu b} ) 
\label{ym34}
\end{eqnarray}
can be carried out by using techniques of homological perturbation theory
and is discussed in Appendix~\ref{appA}.
The result is
\begin{eqnarray}
{\cal H}_1 & = & - \sum_{j=1}^{n-1} 
\rho_1^{(n-j)}\rho_1^{(j)} \Big [ 
{\cal S}_0 ( \int d^4x \, A_\mu^a \frac{\delta}{\delta A_\mu^a} 
{\cal S}_0 ( \int d^4y \, \hat A^{b*'}_\nu A^{\nu b} ))
\nonumber \\
& & 
~~~~~~~~~ - {\cal S}_0 \Big ( \int d^4x \, 
\frac{1}{g^2} ( \square A^d_\rho - \partial_\rho (\partial A)^d ) A^{\rho d} \Big ) \nonumber \\
& & 
~~~~~~~~~ + {\cal S}_0 \Big ( \int d^4x \, 
\frac{2}{g^2} f^{dlm} A^l_\sigma (\partial_\rho A^m_\sigma -
					 \partial_\sigma A^m_\rho) A^d_\rho 
\Big ) \nonumber \\
& & 
~~~~~~~~~ - {\cal S}_0 \Big ( \int d^4x \, 
\frac{1}{4g^2} f^{vqk}f^{krd} A_\sigma^q A_\sigma^r 
A_\rho^v A_\rho^d \Big )
\Big ] \nonumber \\
\label{ym35}
\end{eqnarray}
Therefore the full functional $\tilde \zeta_{\bf 0}^{(n)}$ is
\begin{eqnarray}
\tilde \zeta_{\bf 0}^{(n)} & = & 
- \int d^4x A_\mu^a \frac{\delta}{\delta A_\mu^a} 
\, \left [ 
\sum_{j=1}^{n-1} \rho_1^{(n-j)} \lambda_1^{(j)}  
\int d^4y \, G^{\rho\sigma b} G_{\rho\sigma}^b 
\right ]
\nonumber \\
& & 
- \sum_{j=1}^{n-1} \rho_1^{(n-j)}\rho_1^{(j)} 
\int d^4x \, A_\mu^a \frac{\delta}{\delta A_\mu^a} 
{\cal S}_0 ( \int d^4y \, \hat A^{b*'}_\nu A^{\nu b} )
\nonumber \\
& & 
+ \sum_{j=1}^{n-1} 
\rho_1^{(n-j)}\rho_1^{(j)} \Big (
 \int d^4x \, 
\frac{1}{g^2} ( \square A^d_\rho - \partial_\rho (\partial A)^d ) A^{\rho d}
\nonumber \\
& &
 - \int d^4x \, 
\frac{2}{g^2} f^{dlm} A^l_\sigma (\partial_\rho A^m_\sigma -
					 \partial_\sigma A^m_\rho) A^d_\rho 
 + \int d^4x \, 
\frac{1}{4g^2} f^{vqk}f^{krd} A_\sigma^q A_\sigma^r 
A_\rho^v A_\rho^d 
\Big ) \, .
\nonumber \\
\label{ym36}
\end{eqnarray}
The action-like part of the $n$-th order symmetric quantum effective action 
is then
\begin{eqnarray}
t^4 \GG^{(n)} & = & 
\lambda_1^{(n)} \int d^4x \, G_{\mu\nu}^a G^{\mu\nu}_a 
+ \rho_1^{(n)} {\cal S}_0 (\int d^4x \, \hat A^{a*'}_\mu A^a_\mu ) 
\nonumber \\
& & 
+ \int d^4x A_\mu^a \frac{\delta}{\delta A_\mu^a} 
\, \left [ 
\sum_{j=1}^{n-1} \rho_1^{(n-j)} \lambda_1^{(j)}  
\int d^4y \, G^{\rho\sigma b} G_{\rho\sigma}^b 
\right ]
\nonumber \\
& & 
+ \sum_{j=1}^{n-1} \rho_1^{(n-j)}\rho_1^{(j)} 
\int d^4x \, A_\mu^a \frac{\delta}{\delta A_\mu^a} 
{\cal S}_0 ( \int d^4y \, \hat A^{b*'}_\nu A^{\nu b} )
\nonumber \\
& & 
- \sum_{j=1}^{n-1} 
\rho_1^{(n-j)}\rho_1^{(j)} \Big (
 \int d^4x \, 
\frac{1}{g^2} ( \square A^d_\rho - \partial_\rho (\partial A)^d ) A^{\rho d}
\nonumber \\
& &
- \int d^4x \, 
\frac{2}{g^2} f^{dlm} A^l_\sigma (\partial_\rho A^m_\sigma -
					 \partial_\sigma A^m_\rho) A^d_\rho 
 + \int d^4x \, 
\frac{1}{4g^2} f^{vqk}f^{krd} A_\sigma^q A_\sigma^r 
A_\rho^v A_\rho^d 
\Big ) \, .
\nonumber \\
\label{ym37}
\end{eqnarray}
Notice the appearance of non-symmetric counterterms in $t^4 \GG^{(n)}$,
due to the lower-order contributions in eq.(\ref{ym37}).
The non-symmetric counter-terms, 
depending on the lower order contributions, 
disappear if one chooses to impose the following normalization
conditions for $t^4\GG^{(j)}$
\begin{eqnarray}
\lambda^{(j)}_1=0 \, , ~~~~~ 
\rho_1^{(j)}=0 \, , ~~~~~ 
j=1,2,\dots,n-1
\label{ym38}
\end{eqnarray}
equivalent to
\begin{eqnarray}
\xi^{(j)}_{G^a_{\mu\nu}G^{\mu\nu a}}=0 \, , ~~~~
\xi^{(j)}_{\hat A_\mu^{a*'}\partial^\mu \omega_a} = 0 \, , ~~~~
j=1,2,\dots,n-1 \, .
\label{ym39}
\end{eqnarray}
They extend the one-loop normalization conditions in eq.(\ref{ym21}).

\section{Conclusions}\label{conc}

In this paper we have analyzed the equivalence between the
recursive restoration of the ST identities for anomaly-free
gauge theories and the parameterization of the local
approximation of the symmetric quantum effective action
in terms of suitable ST functionals, associated with the
cohomology classes of the classical linearized ST operator
${\cal S}_0$.

The ST functionals of dimension $\leq 4$ correspond to the invariant
counterterms, those of dimension $>4$ generate the non-symmetric
counterterms upon projection on the action-like sector. 
At orders higher than one in the loop expansion there are additional
contributions to the non-invariant counterterms, arising from
known lower order terms. They can also be parameterized by using
the ST functionals.

We have applied the method to Yang-Mills theory 
in the Landau gauge
with an explicit mass term introduced in a BRST invariant way via
the BRST doublet $(\bar \rho,m)$
and we have derived the most general solution for the action-like part
of the quantum effective action, compatible with the fulfillment
of the ST identities, the ghost equation, the anti-ghost equation
and the $B$-equation, to all orders in the loop expansion.
By exploiting the ST parameterization we have also provided 
the full dependence of the solution on the normalization conditions, that 
have to be supplemented in order to fix
the parameters left free in the symmetric quantum effective action
after the imposition of the ST identities
and of the other relevant symmetries of the model.

The method can applied whenever the classical linearized ST operator
${\cal S}_0$ is nilpotent. The cohomology classes of ${\cal S}_0$
in local jet spaces without power-counting restrictions 
have to be known in order to explicitly work out the form of the
ST functionals and hence to solve the parameterization problem.
Such a cohomology problem can be treated for a large class of theories
of physical interest by applying the powerful 
methods of homological perturbation theory.

\section*{Acknowledgments}

Many useful discussions with R.~Ferrari, W.~Hollik and D.~Maison
are gratefully acknowledged. The author wishes to thank the
Theory Group at the University of Milano, where part of this
work was completed, for the kind hospitality.

\appendix
\section{Appendix}\label{appA}
In this Appendix we analyze the term
\begin{eqnarray}
{\cal H}_1 =  
\sum_{j=1}^{n-1} 
\rho_1^{(n-j)}\rho_1^{(j)} \int d^4x \, (-\partial^\mu \omega^a)
\frac{\delta}{\delta A_\mu^a} {\cal S}_0 (\int d^4y \, 
\hat A^{b*'}_\nu A^{\nu b} ) 
\label{app1}
\end{eqnarray}
contributing to the $n$-th order ST identities and construct a functional
$\tilde \zeta^{(n)}_{{\cal H}_1}$ such that
\begin{eqnarray}
{\cal H}_1 = {\cal S}_0 ( \tilde \zeta^{(n)}_{{\cal H}_1} ) \, .
\label{app2}
\end{eqnarray}
In order to construct  $\tilde \zeta^{(n)}_{{\cal H}_1}$ 
we first notice the following relation:
\begin{eqnarray} 
&& \int d^4x \, \partial_\mu \omega^a \frac{\delta}{\delta A_\mu^a} 
{\cal S}_0 ( \int d^4y \, \hat A^{b*'}_\nu A^{\nu b} )  = 
{\cal S}_0 ( \int d^4x \, A^{\mu}_a \frac{\delta}{\delta A_\mu^a} 
{\cal S}_0 ( \int d^4y \, \hat A^{b*'}_\nu A^{\nu b} )) 
\nonumber \\
&& ~~~~~~~~~~~~~~ + \int d^4x d^4y \, A_\mu^a(x) 
\frac{\delta^2 \G^{(0)}_{m=0}}{\delta A_\mu^a(x) \delta A^d_\rho(y)}
\frac{\delta}{\delta \hat A(y)^{d*'} _\rho}
{\cal S}_0 ( \int d^4z \, \hat A^{b*'}_\nu A^{\nu b} ) \, .
\label{app3}
\end{eqnarray}
The R.H.S. in the first line of the above equation is 
${\cal S}_0$-exact. Therefore the analysis is reduced to the functional
\begin{eqnarray}
\Delta' & = & \int d^4x d^4y \, A_\mu^a(x) 
\frac{\delta^2 \G^{(0)}_{m=0}}{\delta A_\mu^a(x) \delta A^d_\rho(y)}
\frac{\delta}{\delta \hat A(y)^{d*'} _\rho}
{\cal S}_0 ( \int d^4z \, \hat A^{b*'}_\nu A^{\nu b} ) \nonumber \\
& = & - \int d^4x d^4y \, A_\mu^a(x) 
\frac{\delta^2 \G^{(0)}_{m=0}}{\delta A_\mu^a(x) \delta A^d_\rho(y)}
\partial^\rho \omega^d \, .
\label{app4}
\end{eqnarray}
We remark that $\Delta'$ is independent of the external sources and 
satisfies the Wess-Zumino consistency condition
\begin{eqnarray}
{\cal S}_0 (\Delta') = s \Delta' = 0 \, .
\label{app5}
\end{eqnarray}
$\Delta'$ depends only on $A_\mu^a$ and $\omega^a$. In order to construct
a functional $\tilde \zeta'$ such that
\begin{eqnarray}
s \tilde \zeta' = \Delta' 
\label{app6}
\end{eqnarray}
we use the techniques of homological perturbation theory \cite{Henneaux:ig}.
We consider the grading induced by the counting operator for the fields
$A_\mu^a$
\begin{eqnarray}
{\cal N} = \int d^4x \, A_\mu^a \frac{\delta}{\delta A_\mu^a} \, .
\label{app7}
\end{eqnarray}
The restriction of $s$ to the subspace spanned by
$A_\mu^a,\omega^a$ and their derivatives 
decomposes according to this grading into
\begin{eqnarray}
s = s_{-1} + s_0 \, , 
\label{app8}
\end{eqnarray}
where
\begin{eqnarray}
&& s_{-1} A_\mu^a = \partial_\mu \omega_a \, , ~~~~ s_{-1} \omega_a = 0 \, ,
\nonumber \\
&& s_0 A_\mu^a = f^{abc} A_\mu^b \omega^c \, , ~~~~
s_0 \omega^a = -\frac{1}{2} f^{abc} \omega^b \omega^c \, .
\label{app9}
\end{eqnarray}
The differentials $s_{-1},s_0$ satisfy
\begin{eqnarray}
s_0^2 = s_{-1}^2 = 0 \, , ~~~~ \{ s_{-1}, s_0 \} = 0 \, ,
\label{app10}
\end{eqnarray}
as a consequence of the nilpotency of $s$.
We notice that $s_{-1}$ is the Abelian approximation to $s$. 
$s_{-1}$ admits the homotopy operator \cite{Zumino}
\begin{eqnarray}
\!\!\!\!\!\!\!\!\!\!\!\!\!\!
\kappa = \int_0^1 dt \left ( A_\mu^a \lambda_t 
\frac{\partial}{\partial (\partial_\mu \omega^a)} 
+ \partial_{(\mu} A_{\nu )}^a \lambda_t 
\frac{\partial}{\partial (\partial_\mu \partial_\nu \omega^a)}
+ \partial_{(\mu} \partial_\nu A^a_{\sigma)}\lambda_t
\frac{\partial}{\partial(\partial_\mu \partial_\nu \partial_\sigma \omega^a)}
+ \dots \right )
\label{app11}
\end{eqnarray}
where $(\mu,\dots,\sigma)$ denotes total symmetrization and the operator
$\lambda_t$ acts as 
\begin{eqnarray}
&& \lambda_t X(\omega^a, F^a_{\mu\nu},
\partial^\sigma  F^a_{\mu\nu}, \partial^\sigma \omega^a, A^\mu_a,
\partial_{\mu_1} \dots \partial_{\mu_{n-2}} 
\partial_{(\mu_{n-1}} A^a_{\mu_n )})  \nonumber \cr
&& ~~~~~~~~~ 
= X (\omega^a, F^a_{\mu\nu},
\partial^\sigma  F^a_{\mu\nu}, t \partial^\sigma \omega^a, t A^\mu_a,
t \partial_{\mu_1} \dots \partial_{\mu_{n-2}} 
\partial_{(\mu_{n-1}} A^a_{\mu_n )}) 
\label{app11_bis}
\end{eqnarray}
where $F_{\mu\nu}^a$ is 
\begin{eqnarray}
F_{\mu\nu}^a = \partial_\mu A_\nu^a - \partial_\nu A_\mu^a \, .
\label{app11_ter}
\end{eqnarray}
We also expand $\Delta'$ according to the degree induced by ${\cal N}$.
We get
\begin{eqnarray}
\Delta' = \Delta'_1 + \Delta'_2 + \Delta'_3 
\label{app12}
\end{eqnarray}
where
\begin{eqnarray}
\Delta'_1 & = & -\frac{1}{g^2} (\square A^d_\rho 
                               -\partial_\rho (\partial A)^d ) \partial^\rho \omega^d \, , \nonumber \\
\Delta'_2 & = & -\frac{1}{g^2} (4 f^{dlm} A_\sigma^l \partial_\sigma A_\rho^m
+2 f^{dlm} (\partial A)^l A_\rho^m -2 f^{dqk} A_\sigma^q 
\partial_\rho A_\sigma^k ) \partial^\rho \omega^d \, , \nonumber \\
\Delta'_3 & = & \frac{3}{g^2} f^{dqk}f^{krv} 
	                       A^q_\sigma A^r_\sigma A^v_\rho \partial^\rho \omega^d \, .
\label{app13}
\end{eqnarray}
We now prove that it is possible to solve the condition
\begin{eqnarray}
s \tilde \zeta' = \Delta'
\label{app14}
\end{eqnarray}
for $\tilde \zeta'$. Eq.(\ref{app14}) reads at order $j$ in the
grading induced by ${\cal N}$:
\begin{eqnarray}
s_{-1} \tilde \zeta'_{j+1} + s_0 \tilde \zeta'_j = \Delta'_j \,,  ~~~~
 j =0,1,2,3 \, .
\label{app15}
\end{eqnarray}
At order zero eq.(\ref{app15}) is verified by setting
\begin{eqnarray}
\tilde \zeta'_{1}=\tilde \zeta'_0 = 0 \, .
\label{app16}
\end{eqnarray}
We assume that eq.(\ref{app15}) is verified up to order $n-1$.
At order $n$ eq.(\ref{app15}) gives
\begin{eqnarray}
s_{-1} \tilde \zeta'_{n+1} = \Delta'_n - s_0 \tilde \zeta'_n \, .
\label{app17}
\end{eqnarray}
Since $s_{-1}$ is nilpotent, there is a consistency condition
on the R.H.S. of eq.(\ref{app17}):
\begin{eqnarray}
s_{-1} (\Delta'_n - s_0 \tilde \zeta'_n) = 0 \, .
\label{app18}
\end{eqnarray}
The above equation is verified since
\begin{eqnarray}
s_{-1}  (\Delta'_n - s_0 \tilde \zeta'_n) & = &
s_{-1} \Delta'_n + s_0 s_{-1} \tilde \zeta'_n \nonumber \\
& = & s_{-1} \Delta'_n + s_0 (\Delta'_{n-1} - s_0 \tilde \zeta'_{n-1} )
\nonumber \\
& = & s_{-1} \Delta'_n + s_0 \Delta'_{n-1} \nonumber \\
& = & 0 \, .
\label{app19}
\end{eqnarray}
In the second line of the above equation we have used the recursive
condition that eq.(\ref{app15}) is verified at order $n-1$.
In the last line we have used the fact that $\Delta'$ fulfills
eq.(\ref{app5}).

Then we can solve eq.(\ref{app17}) by using the homotopy operator
$\kappa$:
\begin{eqnarray}
\tilde \zeta'_{n+1} = \kappa (
 \Delta'_n - s_0 \tilde \zeta'_n  ) \, .
\label{app20}
\end{eqnarray}
By applying this procedure we get
\begin{eqnarray}
\tilde \zeta' & = &  - \int d^4x \, 
\frac{1}{g^2} ( \square A^d_\rho - \partial_\rho (\partial A)^d ) A^{\rho d}  \nonumber \\
& &  + \int d^4x \, 
\frac{2}{g^2} f^{dlm} A^l_\sigma (\partial_\rho A^m_\sigma -
					 \partial_\sigma A^m_\rho) A^d_\rho 
\nonumber \\
& & 
-  \int d^4x \, 
\frac{1}{4g^2} f^{vqk}f^{krd} A_\sigma^q A_\sigma^r 
A_\rho^v A_\rho^d \, .
\label{app21}
\end{eqnarray}
The final result for ${\cal H}_1$, combining eq.(\ref{app3})
and eq.(\ref{app21}), is given in eq.(\ref{ym35}).

\end{document}